      [1999/12/01 v1.4c Il Nuovo Cimento]
\documentclass{cimento}
\usepackage{graphicx}

\title{Achromatic Breaks for \textit{Swift} GRBs: Any Evidence?}
\author{S.~Covino\from{ins:oab}\ETC,
D.~Malesani\from{ins:dark}\from{ins:sissa},
G.~Tagliaferri\from{ins:oab},
S.D.~Vergani\from{ins:duns}\from{ins:DCU},
G.~Chincarini\from{ins:oab}\from{ins:bic},
D.A.~Kann\from{ins:kan},
A.~Moretti\from{ins:oab},
L.~Stella\from{ins:rome}
on behalf of the MISTICI collaboration}
\instlist{\inst{ins:oab} INAF, Osservatorio Astronomico di Brera, via E. Bianchi 46, I-23807, Merate (LC), Italy
             \inst{ins:dark} Dark Cosmology Centre, Niels Bohr Institute, University of Copenhagen, Juliane Maries vej 30, DK-2100 Copenhagen \O, Denmark
             \inst{ins:sissa} Internationl School for Advanced Studies (SISSA), via Beirut 2-4, I-34014, Trieste, Italy
             \inst{ins:duns} Dunsink Observatory-DIAS, 5 Merrion Square, Dublin 2, Ireland 
	    \inst{ins:DCU} School of Physical Sciences and NCPST, Dublin City University, Dublin 9, Ireland
	    \inst{ins:bic} Dipartimento di Fisica, Universit\`a degli Studi Milano-Bicocca, Piazza delle Scienze 3, I-20216, Milano, Italy
	    \inst{ins:kan} Th\"uringer Landessternwarte Tautenburg,  Sternwarte 5, D-07778 Tautenburg, Germany
             \inst{ins:rome} INAF, Osservatorio Astronomico di Roma, via di Frascati 33, I-00040, Monteporzio Catone (Roma), Italy
}
\PACSes{
\PACSit{98.70.Rz}{gamma-ray sources; gamma-ray bursts}
\PACSit{95.30.Lz}{Hydrodynamics}
}
\begin{document}

\maketitle

\begin{abstract}
The availability of multi-wavelength high-quality data of gamma-ray burst
afterglows in the \textit{Swift} era, contrary to the expectations, did not
allow us to fully confirm yet one of the most fundamental features of the
standard afterglow picture: the presence of an achromatic break in the decaying
light curve. We briefly review the most interesting cases identified so far.
\end{abstract}

\section{Introduction}

In the currently accepted picture, the material powering gamma-ray burst
emission is believed to be collimated in a narrow outflow with opening angle
$\vartheta_{\rm j}$. Following this scenario, the afterglow models predict a
steepening in the light curve decay.
As long as the bulk Lorentz factor is $\Gamma \ge 1/\vartheta_{\rm
j}$, an observer whose line of sight is inside this angle has no knowledge
of what is outside the jet, and the outer edge of the jet visible to the observer 
expands just like in the spherically symmetric case. As the
flow decelerates, the Lorentz factor eventually drops below $1/\vartheta_{\rm
j}$, so that
the observer sees the physical edge of the jet and the
observed light curve decays faster. This break should happen simultaneously at all
frequencies \cite{ref:rho99,ref:sph99}.

Indeed, in observations over a limited wavelength range, transitions involving a temporal
steepening consistent with the theoretical expectations for a jet break have
been seen in many cases \cite{ref:Zeh06} (see for instance the prototypical
case of GRB\,990510 \cite{ref:isra99,ref:sta99,ref:har99}). The jet opening
angle inferred through these observations imply a significant degree of collimation, 
so that the total energy budget emitted by GRBs is close to
$E_\gamma \sim 10^{51}$~erg \cite{ref:fra01}. Moreover, following the
interpretation of these breaks as due to jetted geometry, several correlations
between the beaming-corrected energetics and other GRB properties have been
discovered \cite{ref:ghir06}, reinforcing the interpretation of these
transitions as actual jet breaks. 

Although the behaviour in the asymptotic regimes ({i.e.} well before and
after the transition) is well-known, there is still no full consensus about the
details of the break shape, where many effects might play a role. However,
roughly independently of the jet structure, the break is predicted to be
essentially achromatic apart from minor effects which do not deeply affect the
overall scenario \cite{ref:mesz06,ref:zha04}. It is therefore
important to verify whether full achromaticity of breaks interpreted as due to
jets are confirmed by means of prompt and follow-up observations in
the \textit{Swift} era.

\section{Optical and X-ray observations}

The \textit{Swift} prompt response to GRBs has allowed unprecedented coverage
of the X-ray afterglow evolution from a few tens of seconds up to several
weeks after the high-energy event \cite{ref:geh04}. The coverage in the
optical band, on the contrary, could not regularly be of comparable quality.
This is due to a combinations of factors. First, there are no large
optical facilities on a regular basis devoted to GRB studies. The large increase
in the number of GRBs promptly located with arcsec positional accuracy
(about 2--3 per week following the \textit{Swift} launch)
has more than compensated the sometime generous time allocation at various
observatories. Groups involved in follow-up studies are often forced to
concentrate the efforts on a few events neglecting several others.
Second, continuous observations are more difficult from ground, due to a number
of constraints (visibility, weather conditions, etc.). Moreover,
worldwide coordination among afterglow observers to combine together all the
observations is still lacking. In particular, when only a few datapoints are
available, data frequently remain unpublished or poorly calibrated.
Last, as a matter of fact, the optical counterparts of the GRBs localised by
\textit{Swift} are also on average fainter than those localised by previous
missions like \textit{Beppo}SAX and HETE-II \cite{ref:rom06}, probably
due to an average larger redshift \cite{ref:jakob06}. This makes it difficult
for medium-sized telescopes to follow the afterglow decay long enough to
collect a complete light curve. The UVOT telescope \cite{ref:rom05} onboard
\textit{Swift} is seldom able to monitor the optical afterglow evolution
for longer than a few hours. 

To further complicate the picture, it is now clear that afterglow light curves
are much more rich than previously thought, displaying
rebrightenings, flares, phases of shallow and steep decay 
\cite{ref:taglia05,ref:chinc06,ref:burrows05,ref:nousek06}. At least some of these 
behaviours are due to extra components contributing to the flux (possibly originating 
from prolonged activity in the GRB central engine), and to a complex jet dynamical evolution
(which may not be adiabatic in the first hours after the GRB). Some of the
mechanisms shaping the light curves are also hydrodynamical, and can produce
achromatic breaks not related to geometric effects. 

Despite the above
limitations, it is quite surprising that among the 180 bursts so far detected by
\textit{Swift}, Fall of 2006, only a few showed clear breaks in their optical light
curves. In many cases, \textit{no} break at all could be seen in the optical,
despite extensive monitoring. Therefore, just a handful of cases are left to
evaluate whether their breaks are achromatic across the optical and X-ray
bands. In the following we present a few examples of combined optical/X-ray
light curves, and try to assess whether their breaks can be due to the jet
effect.

\section{Test cases}

Among the \textit{Swift} GRBs for which a detailed light curve is
available in the literature, the best candidates to look for achromatic
breaks are so far: GRB\,050525A, GRB\,050801, GRB\,060124 and
GRB\,060526. For all these events we could single out a possible achromatic
break, while the light curve coverage is adequate for a comprehensive discussion.

\subsection{GRB\,050525A}

\begin{figure}
\begin{tabular}{cc}
{\centering \includegraphics[width=6.5cm]{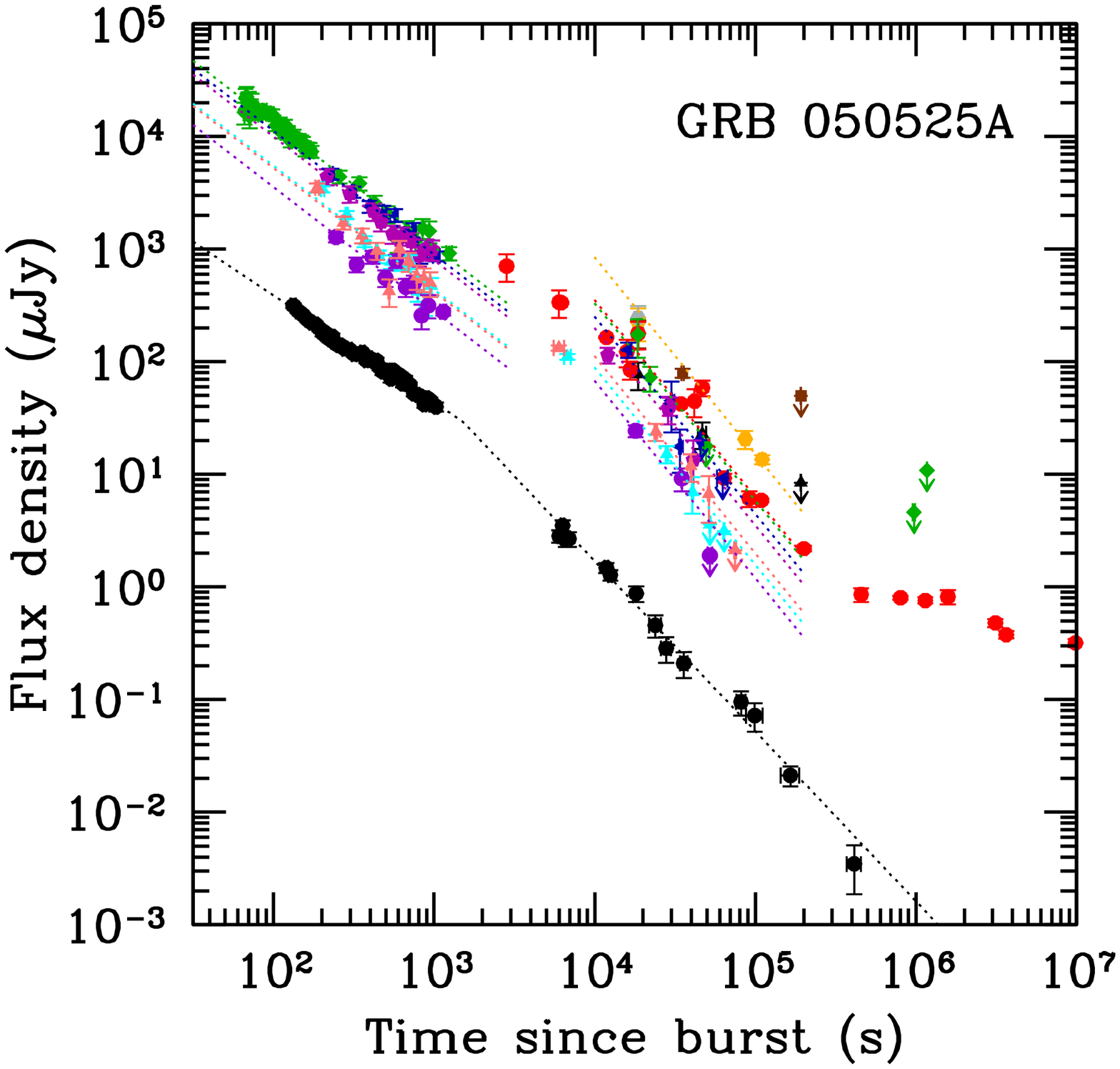}} &
{\centering \includegraphics[width=6.5cm]{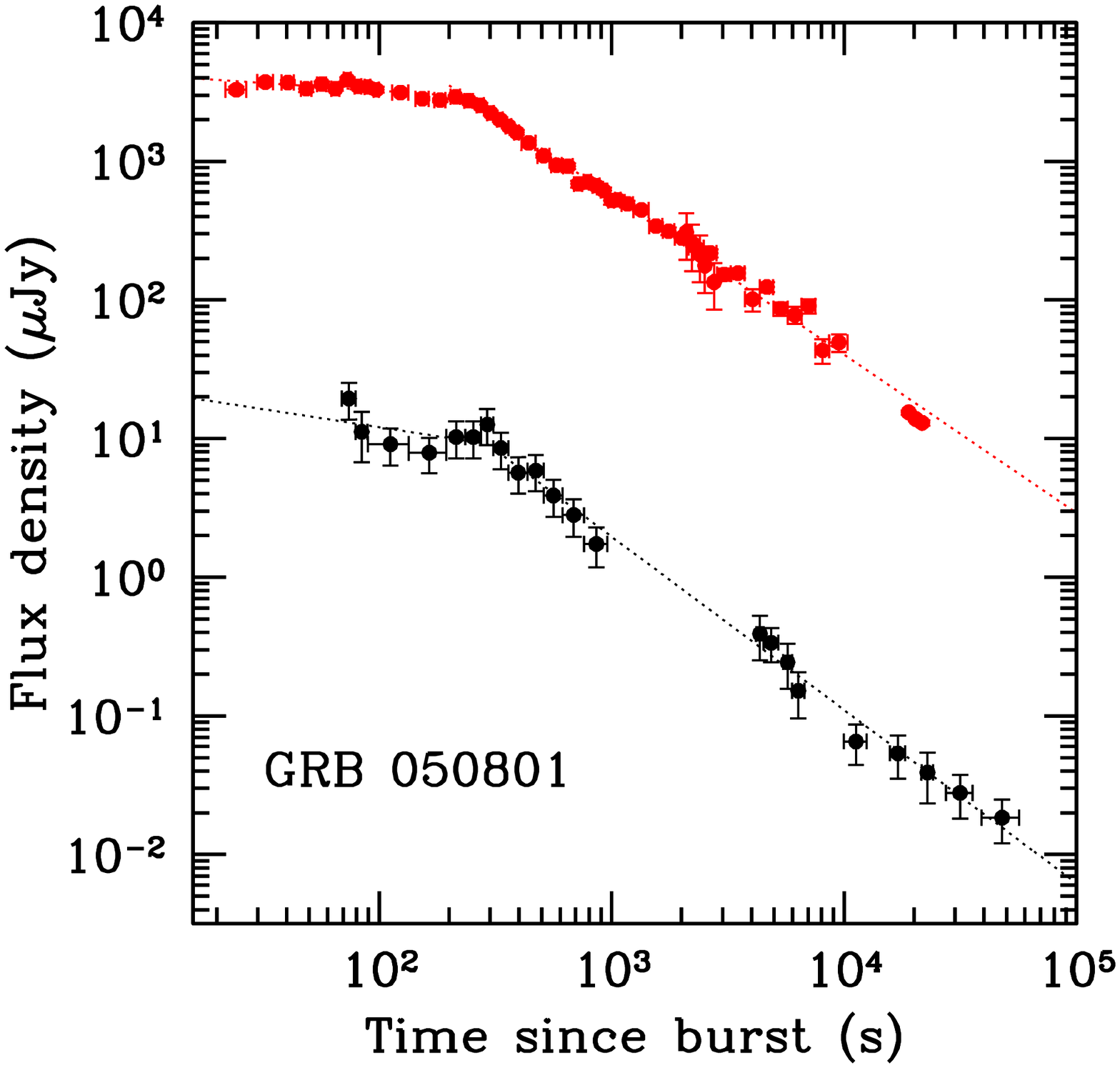}}
\end{tabular} 
\caption{Optical and X-ray light curves of GRB\,050525A (left) and GRB\,050801 (right).}
\label{fig:12}
\end{figure}

GRB\,050525A is one of the best sampled GRBs detected by \textit{Swift}, 
at both optical and X-ray wavelengths \cite{ref:blu05,ref:mdv06}.
The light curve of GRB\,050525A presents some deviations with respect to
a simple broken power law (see Fig.~\ref{fig:12}, left), both in the optical
and X-rays. The initial afterglow was indeed modeled by including also a
contribution from the reverse shock \cite{ref:sha05,ref:blu05}. The presence of
an achromatic break is still being discussed. Blustin \etal{} \cite{ref:blu05}
identify a break at $\sim 0.15$ days, a time which is consistent
with the break being simultaneous in the optical and X-ray bands. On the other hand,
Della Valle \etal{} \cite{ref:mdv06} find a break in the $R$ band at a slightly
later time of $t \sim 0.3$~days. \textit{Swift}-XRT and -UVOT data before the
break show a decay and spectral slope consistent with the predictions of
standard afterglow models for a uniform interstellar medium \cite{ref:sari98}.
Moreover there is no spectral evolution before and after the break. However the temporal index
of the post-break decay appears to be too shallow ($\alpha_{\rm post} =
1.6$--1.8) compared to the predictions ($\alpha_{\rm post} = p$, where $p
\approx 2.2$ is the electron energy distribution index). Apart from the low
inferred $p$, it is possible that the decay is shallower due to inefficient
sideways expansion \cite{ref:panmes99}. However, given the uncertainty in the
late decay indices (also given the contribution from the associated SN\,2005nc
\cite{ref:mdv06}), it is also possible to model the light curves with a steeper
post-break decay, assuming that the jet break takes a finite time to
complete and the post-break asymptotic regime has not yet been reached.
Recently Sato \etal{} \cite{ref:sato06} questioned the identification
of an achromatic break for GRB\,050525A, due to the lack of agreement between
broad-band modeling of the afterglow light curves around the break and the
standard afterglow model predictions. The issue is therefore still to be
settled. GRB\,050525A might be one of the strongest outliers for the so
called Ghirlanda et al. relation \cite{ref:ghir04}. However, it should be also
stressed that while the presence of an achromatic break is essentially due to
the outflow geometry only, the spectral and temporal indices depend on
more subtle and model-dependent details. The collimation factor, and thus
the real energetics of the burst, are likely not much affected by these
details.

\subsection{GRB\,050801}

The optical afterglow of GRB\,050801 was detected already $\sim 22$\,s after
the burst \cite{ref:ryk06} and the light curve was then followed for more than
10,000~s (see Fig.~\ref{fig:12}, right). The combined \textit{Swift}-XRT and
optical light curves identify a clear achromatic break as early as
$\sim 250$\,s. Indeed, the whole afterglow evolution did not show any sign of
spectral evolution between optical and X-ray, and a single power-law can
account for the broad-band spectrum. In spite of the convincingly achromatic nature
of this break, interpretation of this transition in term of a jet break is
difficult. The spectral and temporal power-law indices are compatible with an
outflow moving in a constant density ISM, and rule out a wind environment. The
post-break decay is  shallow, $\alpha \sim 1.3$, requiring therefore a value
for the electron energy distribution index as low as $p \sim 1.3$. Such an
extreme value is however inconsistent with the essentially flat afterglow evolution observed
before the break. A jet break already $\sim 250$\,s after the
burst would be the earliest detected. An alternative and intriguing
possibility is that the achromatic break is due in fact to the afterglow onset
\cite{ref:ryk06,ref:kozh06,ref:cenk06,ref:mol06}.

\subsection{GRB\,060124}

\begin{figure}
\begin{tabular}{cc}
{\centering \includegraphics[width=6.5cm]{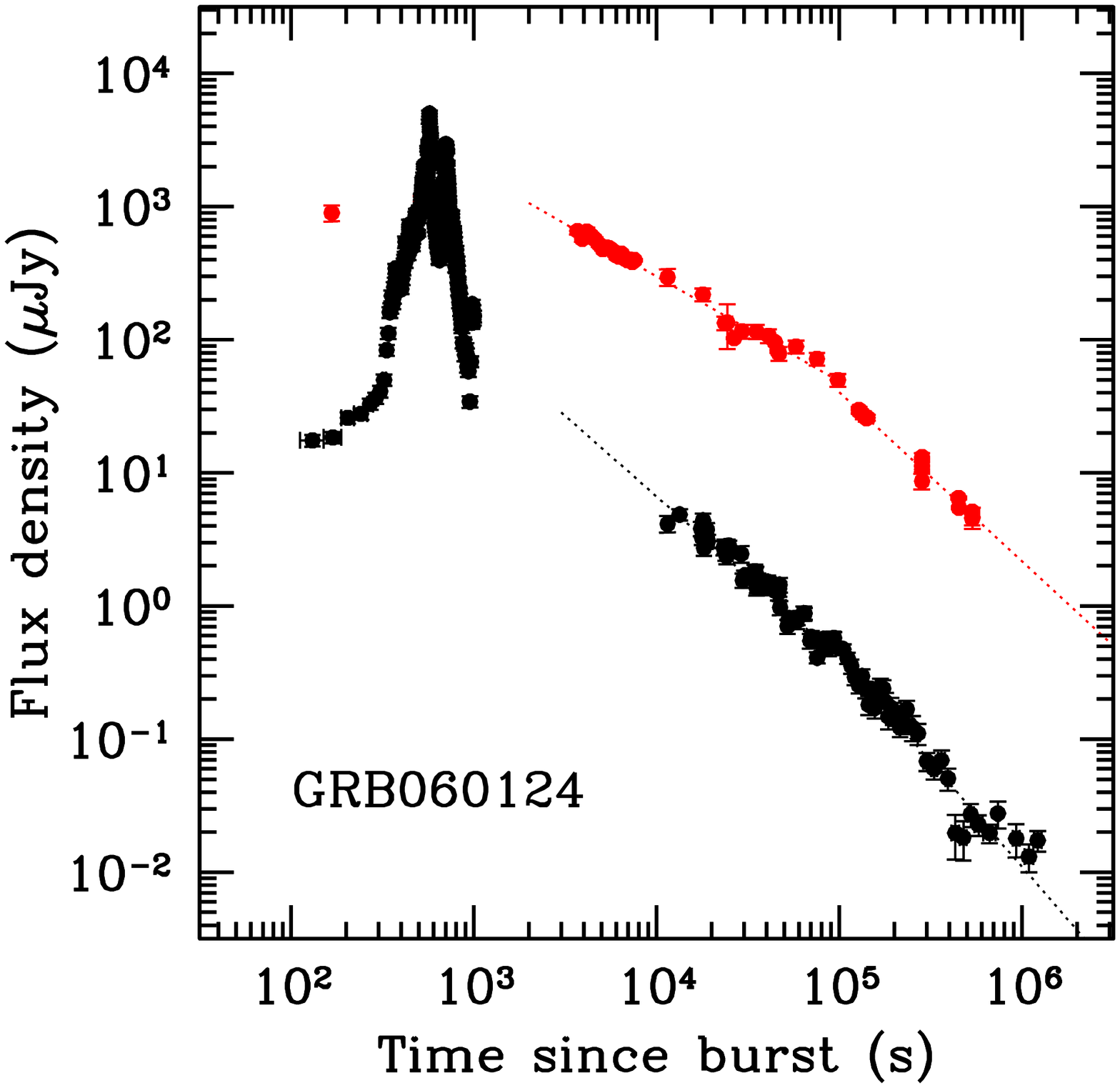}} &
{\centering \includegraphics[width=6.5cm]{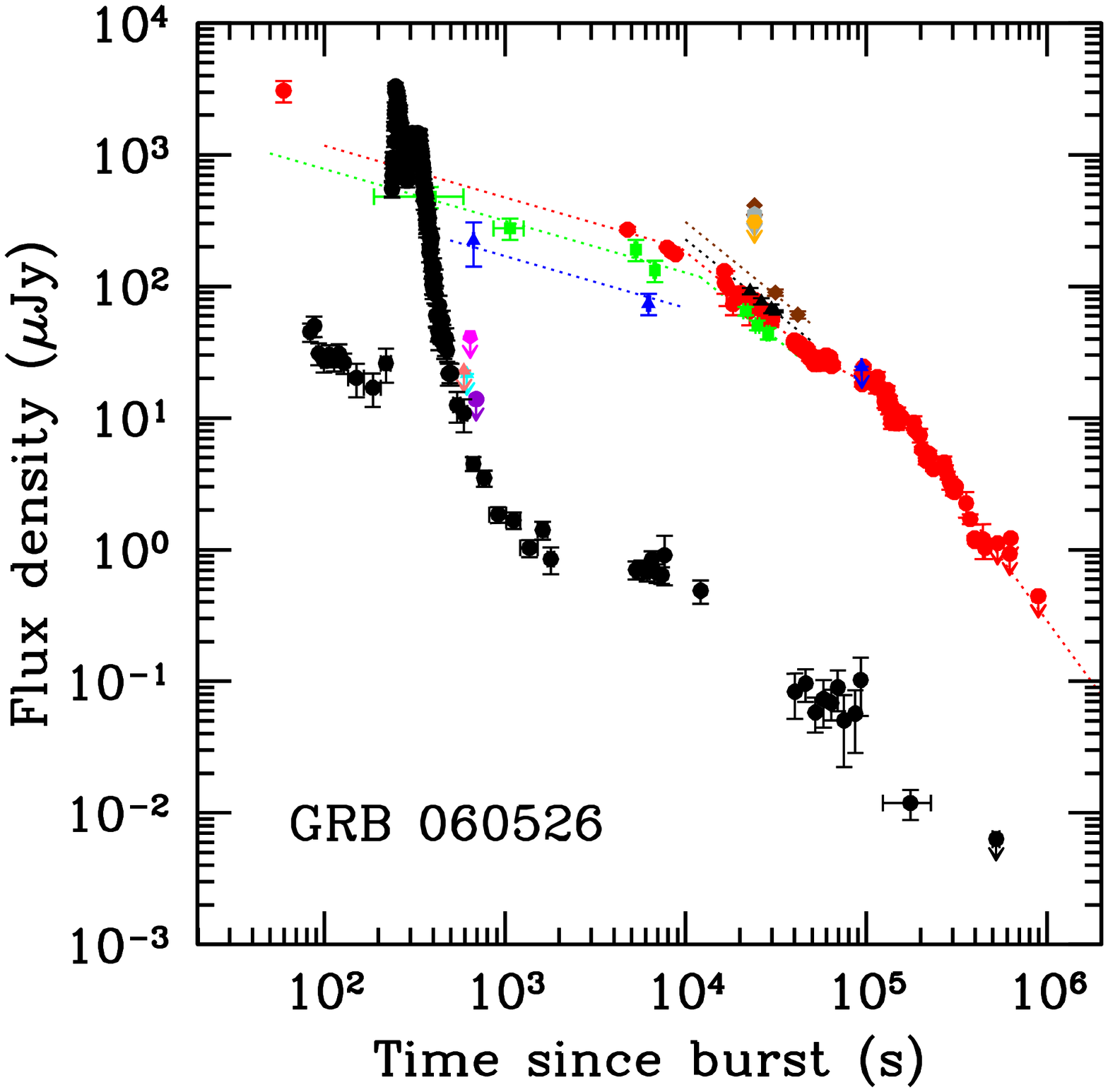}}
\end{tabular}
\caption{Optical and X-ray light curves of GRB\,060124 (left) and GRB\,060526 (right). For GRB\,060124 optical data were taken in different filters and reported to the $R$ band adopting the average spectrum \cite{ref:curr06,ref:kan06}}
\label{fig:34}
\end{figure}

The most striking example of achromatic break in this sample comes from the
analysis of the light curve of GRB\,060124 (see Fig.~\ref{fig:34}, left). The
optical and X-ray light curves were well sampled before and after the
achromatic break identified at $t_{\rm b} \sim 1$~day \cite{ref:curr06,ref:kan06}. The
pre-break decay ($\alpha_{\rm X} \sim 1.1$, $\alpha_{\rm opt} \sim 0.8$) and
spectral indices ($\beta_{\rm X} \sim 1.0$, $\beta_{\rm opt} \sim 0.4$) are in
good agreement with the expectations of the standard fireball theory in the
case of homogenous ISM and with an electron energy distribution index $p \sim
2$. However, problems with the identification of this transition as a jet break
arise because the post-break temporal decay indices are both too shallow and
different in the X-ray and optical ($\alpha_{\rm X} \sim 1.7$, $\alpha_{\rm
opt} \sim 1.3$). Shallower post-break decays might be produced if, for some
reason, the jets are not spreading as effectively as expected. Further
hypotheses are required to account for the different decay indices that, in the
context of constant density ISM, should equal the electron energy distribution
index independent of the wavelength.

\subsection{GRB\,060526}

The X-ray and optical light curves of GRB\,060526 are complex and
present multiple flares and breaks \cite{ref:dai06}. The late-time optical
light curve shows a well-defined steepening (though the
presence of flares may introduce systematic errors in the determination of the
decay index; see Fig.\,\ref{fig:34}, right). The X-ray data, despite their sparse sampling, 
appear to support the presence of such a break, but are also
consistent with an uninterrupted decay. Therefore, an achromatic break is not
strictly required. The pre-break decay index is $\alpha_{\rm pre} \sim 1.1$
while the post-break decay index is $\alpha_{\rm post} \sim 3.4$ (with a large
error), mainly constrained by the optical data. The broad-band modeling of the
late afterglow is consistent with a single synchrotron component, even though
the X-ray data do not strongly constrain the fit. In any case, the
post-break temporal index is too steep to be accounted for in the standard model,
possibly requiring varying micro-physical parameters.

\section{Conclusions}

The intrinsic (and to some extent unexpected) complexity of the \textit{Swift} 
afterglows has not prevented the identification of breaks in many X-ray light curves, that are
sometimes consistent with the requirements of the standard afterglow theory for
jet breaks. Things change considerably once optical data are taken into
account. For only a few events the optical coverage is as good as in the X-rays. 
Although jet breaks are part of the so-called canonical
\textit{Swift}-XRT light curve, achromatic breaks have been identified
only for a handful of events. Furthermore, in no case the jet-break
interpretation holds without the need to introduce additional ingredients. On the
contrary, for a few other events, \textit{chromatic} breaks have also been
identified. Modeling these events within the standard afterglow theory requires
extra assumptions such as a variation of microphysical parameters for the electron and magnetic
energies during the afterglow evolution or, alternatively, that the X-ray and
the optical afterglows arise from different components
\cite{ref:pana06a,ref:pana06b}. 

The paucity of identified jet breaks over a large wavelength range can of course
affect the interpretation of correlations such as the Ghirlanda et al. relation
\cite{ref:ghir04}. It should be stressed that these relations are so far
derived mainly (or only) by means of jet breaks identified at optical
wavelengths. It is therefore possible to wonder whether jet breaks identified in the
X-rays only carry the same information about the total energetics of GRBs.
In any case, the relatively
limited energy range covered by BAT onboard \textit{Swift} did not allow in
most cases to measure the spectral peak energy of the prompt GRB emission, thus
preventing the check (and possibly the improvement) of the Ghirlanda et al. relation
with these events, even if a jet break is identified with the XRT. It
should be noted, however, that a model-independent version of the
Ghirlanda relation exists \cite{ref:LZ05}, which involves only the break time (as
measured \textit{in the optical}) and which does not necessarily involve a
geometrical interpretation. This relation may remain valid even in the absence of
an achromatic break, in this case requiring a completely different
interpretation. In any case, a better optical coverage (possibly aided by a
better coordination among observers) could improve the light curve sampling of
\textit{Swift} GRBs and allow a firmer identification of jet break transitions.


\end{document}